\documentstyle[11pt,aaspp4]{article}

\received{8 April 1996}
\accepted{27 May 1996}

\begin{document}     

\newcommand{\squig}{$\sim$}
\newcommand{\decsec}[2]{$#1\mbox{$''\mskip-7.6mu.\,$}#2$}
\newcommand{\decsectim}[2]{$#1\mbox{$^{\rm s}\mskip-6.3mu.\,$}#2$}
\newcommand{\decmin}[2]{$#1\mbox{$'\mskip-5.6mu.$}#2$}
\newcommand{\asecbyasec}[2]{#1$''\times$#2$''$}
\newcommand{\aminbyamin}[2]{#1$'\times$#2$'$}

\title{{\it HST} Imaging of Bright Galactic\\
       X-ray Binaries in Crowded Fields
\footnote{\ Based on observations with the NASA/ESA Hubble Space
Telescope obtained at the Space Telescope Science Institute, which is
operated by the Association of Universities for Research in Astronomy,
Inc., under NASA contract NAS5-26555.}
}
\author{Eric W. Deutsch, Bruce Margon, Stefanie Wachter and Scott F. Anderson }
\affil{Department of Astronomy\\
       University of Washington\\
       Box 351580\\
       Seattle, WA 98195-1580\\
       Electronic mail: deutsch, margon, wachter \& anderson@astro.washington.edu}

\begin{center}
To appear in ApJ Volume 471: November 10, 1996\\
received: 8 April 1996; accepted: 27 May 1996
\end{center}

\begin{abstract}

We report high spatial resolution {\it HST} imagery and photometry of
three well-studied, intense galactic X-ray binaries, X2129+47, CAL 87
and GX17+2. All three sources exhibit important anomalies, not readily
interpreted by conventional models. Each source also lies in a severely
crowded field, and in all cases the anomalies would be removed if much
of the light observed from the ground in fact came from a nearby,
thus-far-unresolved superposed companion. For V1727~Cyg (X2129+470), we
find no such companion.  We also present an {\it HST} FOS spectrum and
broadband photometry which is consistent with a single, normal star.
The supersoft LMC X-ray source CAL~87 was already known from
ground-based work to have a companion separated by \decsec{0}{9} from
the optical counterpart; our {\it HST} images clearly resolve these
objects, and yield the discovery of an even closer, somewhat fainter
additional companion.  Our photometry indicates that contamination is
not severe outside eclipse, where the companions only contribute 20\%
of the light in V, but during eclipse more than half of the V light
comes from the companions.  The previously-determined spectral type of
the CAL~87 secondary may need to be reevaluated due to this significant
contamination, with consequences on inferences of the mass of the
components.  We find no companions to NP~Ser (=X1813--14, =GX17+2).
However, for this object we point out a small but possibly significant
astrometric discrepancy between the position of the optical object and
that of the radio source which is the basis for the identification.
This discrepancy needs to be clarified.
\end{abstract}


\section{INTRODUCTION}

We discuss the optical identification of three intense, well
studied X-ray sources whose optical counterparts remain frustratingly
ambiguous, despite literally decades of work.  The three sources are
members of a very interesting subset of X-ray binary
systems: the low mass X-ray binary (LMXRB). In these systems, the large
intrinsic $L_x/L_{opt}$ ratio often creates spectacular heating effects
on the low-mass normal star, resulting in dramatic photometric and
spectroscopic variability as a function of binary phase.

The three sources discussed here have in common the fact that all
have suggested optical counterparts that have been well studied
spectroscopically and photometrically. However, in all cases these studies
have uncovered anomalies that make the identifications suspect,
contradictory, or impossible to understand in the current framework. All
three objects are also in very crowded fields, two because they are at very
low galactic latitude ($b=1^\circ$ and $b=-3^\circ$), and the third due to
membership in the LMC. The combination of the odd optical properties plus the
severe crowding has led to multiple suggestions over the years that, in each
case, the wrong optical counterpart may have been selected due to the
superposition of an additional object of angular separation too small to be
resolvable from the ground. The superposition could be foreground/background
(the classical ``optical double" of unrelated stars), or perhaps even a
kinematically-related additional object; for the purpose of this discussion
the distinction is unimportant. We have obtained multicolor WFPC2
images of all three objects, to conduct the most sensitive possible probe of
the presence of a superposed image, possibly resolving not only the star, but
the interpretive contradictions.

The suggestion that superpositions may confuse the analysis of very
well-studied galactic X-ray sources is not merely mischievous or a
desperate hope: it has already been demonstrated to occur in at least
one famous case.  The exceptionally intense X-ray source Circinus X-1
(X1516-569) was optically identified almost 20 years ago (Whelan et al.
1977) with an $R=16$ object, and the identification has been presumed
correct since then, based on arcsecond accuracy X-ray and radio
positions, as well as very strong $H\alpha$ emission. However, it has
recently been demonstrated (Moneti 1992, Duncan et al. 1993) that the
``identification" is in fact the wrong star, by more than $1''$! The
field proves to have 3 objects within \decsec{1}{5}. Had the grouping been
slightly more compact than this, only {\it HST} imaging would be capable of
revealing the true nature of the problem. Strangely enough, although
Cir~X-1 is far from fully understood, no interpretive inconsistencies
pointed towards a misidentification (although Argue \& Sullivan 1982
and Argue et al.  1984 indeed questioned the identification on
astrometric grounds). The three objects studied here, each of which
will now be discussed in turn, all do have such existing observational
problems, and thus in some sense are even better candidates for
superposition than the one object where it is already definitely known
to be a problem.

\section{OBSERVATIONS AND DATA REDUCTION}

\subsection{Planetary Camera Imagery}

On July 12 and 13, 1995, we obtained (post-refurbishment mission) {\it
HST} Planetary Camera (PC) images of V1727~Cyg, CAL~87, and NP~Ser.
For each target, exposures through the F336W (2~$\times$~300 s), F439W
(2~$\times$~200 s), F555W (3~$\times$~40 s) and F675W (3~$\times$~50 s)
filters were taken; these filters are similar to Johnson U, B, V, and R
respectively.  The PC frames have been processed through the standard
data reduction pipeline at STScI.  Further reduction was performed with
software written in IDL by E.W.D. or available in the IDL {\it
Astronomy User's Library} (Landsman 1993).

First, each of the sets of two or three exposures were combined with a
cosmic-ray rejection algorithm.  Although each of these fields is
crowded in ground-based images, the PC resolution is such that all
sources are well-separated, so we use aperture photometry to measure
magnitudes.  Aperture corrections are taken from Table 2(a) in Holtzman
et al. (1995a).  The photometric measurements have not been corrected
for geometric distortions in the PC, but the simple correction for
charge transfer efficiency losses detailed in Holtzman et al. (1995a)
has been applied.  Instrumental magnitudes are converted to Johnson
UBVR magnitudes using the transformations presented in equation (8) and
Table 7 in Holtzman et al. (1995b), where transformation errors are
reported to be less than 2\%, except in the F336W filter and for
sources with unusual colors.  The final Johnson magnitudes are the
result of a best fit to the 12 transformation equations.  Magnitudes
and colors for all objects are presented in Table 1.  Statistical and
read noise uncertainties are 1\% or less except for the faint CAL~87
Companions A and B; for these two objects, measurement errors are
reported in the table as well.  Additional systematic errors for all
magnitudes due to uncertainties in detector performance, absolute
calibration and filter transformations are \squig 5\%.

\subsection{Faint Object Spectrograph Data}

On October 1, 1993 we obtained (pre-servicing mission) {\it HST} FOS UV
spectra of V1727~Cyg, consisting of two 600~s exposures through the
G160L grating and two 600~s exposures through the G270H grating, as
part of the FOS Guaranteed Time Observations program.  The
\decsec{4}{3} $\times$ \decsec{4}{3} single aperture was used for all
exposures.  The raw data were reprocessed with the latest CALFOS
reduction algorithm (Lindler \& Bohlin 1994) and calibration files;
each pair of exposures was combined.

UV spectra of red objects with the FOS are contaminated with some
scattered light from longer wavelengths by the diffraction gratings
(e.g., Kinney 1993; Koratkar 1995).  A simple scattered light
subtraction algorithm (Kinney \& Bohlin 1993) is applied to the G160L
spectra as part of the CALFOS reduction; the correction is calculated
from a region on the detector that is not illuminated by dispersed
light.  For the G270H spectra, no CALFOS correction is available, as
light is dispersed over the entire detector.

After the scattered light correction of 0.0020 and 0.0034 s$^{-1}$
pixel$^{-1}$ to the two G160L spectra, the average count rate in the
combined spectrum slowly increases from 0.0015 to 0.0025 s$^{-1}$
pixel$^{-1}$ between 1400 and 2200 \AA.  It is likely
that these residual counts come from a slight gradient in the scattered
light rather than dispersed light from the target.  We set upper
limits of $3\times 10^{-17}$ erg cm$^{-2}$ s$^{-1}$ \AA$^{-1}$ at
1400 \AA\ and $1\times 10^{-17}$ erg cm$^{-2}$ s$^{-1}$ \AA$^{-1}$ at
2200 \AA.

Since there is no CALFOS scattered light correction available for the
G270H grating, which covers the 2250--3300 \AA\ region, the correction
must be determined empirically.  Matching flux levels in the overlap
region between the G160L and the G270H gratings is not possible because
of the non-detection in G160L.  Shortward of 2400 \AA, the count rate
is nearly constant.  After correcting for the instrumental response we
find that the flux level at the shortest wavelengths is well above the
upper limit from the shorter wavelength grating.  This indicates that
the flux below 2400 \AA\ is due to scattered light, so we correct the
spectrum by subtracting the count rate 0.004 s$^{-1}$ pixel$^{-1}$
seen below 2400 \AA.

\section{DISCUSSION OF INDIVIDUAL OBJECTS}

\subsection{V1727 Cyg}

The intense LMXRB X2129+470 is of special interest because it shares
many properties in common with Her~X-1/HZ~Her, a very rare but
important system.  The optical counterpart, V1727~Cyg, at times
exhibits large amplitude ($B\sim1.5$~mag) variations with a period of
5.2~hr (Thorstensen et al.  1979), the orbital period of the system.
The detailed features of the light curve are similar to those of the HZ
Her/Her X-1 system; consequently, the variations are thought to be
caused by heating due to X-ray irradiation of the companion's
atmosphere (McClintock et al. 1981).  Alternatively, it has been
suggested that disk aspect variations are responsible (Chevalier
1989).  Early spectra show weak HeII $\lambda$4686 and CIII/NIII
$\lambda\lambda$4640/50 emission (Thorstensen et al. 1979), which is
typical for a LMXRB. Radial velocity studies (Thorstensen \& Charles
1982; Horne et al. 1986) imply masses of 0.4 $\pm$ 0.2 $M_{\odot}$ for
the companion and 0.6 $\pm$ 0.2 $M_{\odot}$ for the compact star, an
unusually low mass for a neutron star. The X-ray luminosity of the
system is also fairly low for a LMXRB, but an X-ray burst found in
archival X-ray data (Garcia \& Grindlay 1987) confirms that the system
contains a neutron star.

The system was thought to be fairly well understood until Pietsch et
al. (1986) reported that V1727~Cyg had entered a low state in 1983: the
X-ray flux and the large photometric variations vanished.  ROSAT
observations in the quiescent state (Garcia \& Callanan 1995) show that
X-ray eclipses continue even in the low flux state.  The extended
interval of X-ray quiescence prompted a flurry of observations to
determine the ellipsoidal variations of the secondary without the
disturbing effect of X-ray heating in order to verify the anomalous low
neutron star mass.  Surprisingly, {\it neither photometric variations
nor radial velocity variations could be detected} (Thorstensen et al.
1988; Kaluzny 1988; Chevalier et al. 1989; Garcia et al. 1989; Cowley
\& Schmidtke 1990).  The star that is visible in the low state optical
spectrum appears to be a normal F8 IV star (Cowley \& Schmidtke 1990),
although Kaluzny (1988) finds that the colors of the object in the low
state are incompatible with colors of an ordinary star. It has been
variously proposed that this is due to superposition of a
foreground/background star or that X2129+47 is a triple system
(Garcia et al. 1989).

If the system is a triple or superposed by a foreground star, then it
is possible that all previous optical data taken in the high state
are severely contaminated. Contamination by a companion might also
explain the anomalous mass inferred for the secondary. Chevalier et al.
(1989) present a ground-based image taken in good seeing, motivated by
a search for a superposed interloper, and find nothing remarkable.

In our PC observations, the angular resolution is better by an order of
magnitude than feasible from the ground, yet V1727~Cyg remains an
unresolved
single source; there are no significant deviations from the typical
WFPC2 PSF with a FWHM of \decsec{0}{065}.  By subtracting a ``Tiny TIM''
model PSF (Krist 1993), we set upper limits on possible polluting
sources as follows:  for $r>$\decsec{0}{4}, $R>23.5$;
\decsec{0}{15}$>r>$\decsec{0}{4}, $R>22.5$;
\decsec{0}{04}$>r>$\decsec{0}{15}, $\Delta R>27\times r$ where $\Delta
R$ is the magnitude difference between V1727~Cyg and a possible
polluting source.  There are no additional sources detected near
V1727~Cyg down to R=23.5 which have not already been observed by
Chevalier et al. (1989) (although pairs B,C and E,F shown in that paper
are easily resolved).  Our photometry for V1727~Cyg, presented in Table
1, was measured between photometric phase 0.58 and 0.70, according to the
ephemeris of McClintock et al. (1982) (derived when this system was
still optically eclipsing.)  Our UBVR colors are consistent with an F8 type star
with E(B--V)\squig 0.3--0.4 (best fit E(B-V)=0.34).  We find no evidence
of U excess mentioned by Kaluzny (1988).

Our {\it HST} spectrum, although obtained during the X-ray low state
and thus weakly exposed, provides constraints on the possibility of
light from more than just an F8 IV star.  After the initial processing
described in \S 2.2, we increase the signal-to-noise by rebinning the
spectrum to \squig 6 \AA, approximately the resolution of the IUE.  We
then obtained archival IUE spectra of two stars which have similar
spectral types to V1727~Cyg, as reported by Cowley \& Schmidtke (1990)
from their optical spectrum: LWR 15500 of HD 82328 (F6 IV, V=3.17) and
LWR 15402 of HD 90839 (F8 V, V=4.83).

In order to examine a possible difference between the {\it HST} spectrum of
V1727~Cyg and normal stars of similar spectral type, the IUE spectra
are reddened with the extinction curve of Savage \& Mathis (1979), and
then the flux level is scaled by the difference between the V magnitude
of V1727~Cyg and the reddened V magnitude of the IUE object.  We find
that a reddening of E(B--V)=0.40 yields an excellent match.  A previous
estimate of the reddening by Cowley \& Schmidtke (1990) from BV photometry
yielded E(B--V)\squig 0.3.  The {\it HST} spectrum and the reddened,
scaled LWR 15500 spectrum are plotted together in Fig. 1; the other IUE
spectrum matches LWR 15500 very closely and is omitted from the plot.
Therefore, if the reddening of V1727~Cyg is E(B--V)\squig 0.3--0.4, the UV
spectral flux and the UBVR colors are consistent with a single F8 star.
The M V star discussed by Garcia et al. (1989) as a hypothetical third
member of a triple system is many orders of magnitude too faint to
alter the spectrum of the F8 IV star, and the system is far too small
to be spatially resolvable in our data.

\subsection{CAL 87}

CAL 87 (Long et al. 1981) is one of only a small number of known LMXRBs
in the LMC, and also one of the supersoft X-ray sources, a recently
discovered class of X-ray object.  It has been pointed out that
these extremely X-ray luminous (\squig 10$^{38}$ erg s$^{-1}$) objects
may populate our Galaxy in numbers comparable to the more familiar,
Eddington-limited 1-10 keV X-ray binaries, but are rarely detected due
to severe interstellar photoelectric opacity (van den Heuvel et al.
1992, Yungelson et al. 1996).  Although many workers view the compact
object in these systems as quite likely an accreting white dwarf (e.g.
van den Heuvel et al. 1992; Motch et al. 1994; van Teeseling et al.
1996), Cowley et al.  (1990) have presented evidence that CAL~87
contains a black hole, and Hughes (1994) suggests a neutron star
companion for the SMC soft-source RX J0059.2--7138, so the situation is
unclear.

The optical counterpart of CAL~87, discovered by Pakull et al. (1987),
exhibits large photometric variations ($\Delta V$$\sim$$1.2$~mag) with
an orbital period of 10.6~hr (Callanan et al. 1989, Cowley et al.
1990).  CAL~87 appears to be an eclipsing system with a deep primary
minimum and a shallow secondary minimum. Models of the shape of the
lightcurve agree well with the eclipse of an extended disk structure,
and indicate that the disk is the dominant light source in the system.

The optical spectrum of CAL 87 shows He II $\lambda$4686 and $H\alpha$
in emission; however CIII/NIII $\lambda\lambda$4640/50, which is
usually seen in LMXRBs, is absent (Pakull et al. 1988). At minimum
light, CaII H \& K and the G band are seen in absorption; these have
been presumed to arise in the secondary and to indicate a spectral type
of late F (Cowley et al. 1990). Cowley et al. (1990) measured the
radial velocity curve for the He II $\lambda$4686 line, which
presumably arises in the accretion disk.  These velocities, plus the
inferred spectral type of the secondary, lead to the suggestion that
the X-ray source is a black hole.

CAL 87 lies in a very crowded region: Cowley at al. (1991) note a close
optical companion only \decsec{0}{9} away. Any photometry of CAL 87 is
therefore contaminated unless taken in subarcsecond seeing.  Cowley et
al.  (1991) attempted to detect radial velocity variations of the
secondary in the infrared Ca triplet, but find none. They
conclude that the spectrum is most likely dominated by the nearby field
star and that much better spatial resolution will be necessary to
separate the two stars.  Small aperture {\it HST} UV-spectroscopy of
the object by Hutchings et al. (1995) presumably does better isolate
the X-ray source, but of course yields no information on the nature of
contamination of ground-based photometry and spectroscopy by nearby
objects.

Our PC observations easily resolve this companion (which we
designate A) and reveal another, fainter and closer, optical companion
(B).  Figure 2 shows the three objects and the surrounding
\asecbyasec{7}{7} region in the F675W filter, with a limiting
magnitude $R\sim23.5$.  Magnitudes for these sources are listed in
Table 1.  The position angles and distances from CAL~87 for Companions
A and B are 335$^\circ$, \decsec{0}{88} and 210$^\circ$, \decsec{0}{65}
respectively.  CAL~87 itself appears to be a single object, with no
significant deviations from the typical WFPC2 PSF with a FWHM of
\decsec{0}{077}.

Using the photometry of all sources in the PC frame, 4 color-magnitude
diagrams are presented in Fig. 3; objects mentioned in this discussion
are labeled on the diagrams.  The M$_{\rm V}$ scale assumes (m--M)=18.4
and A$_{\rm V}\sim 0.2$.  The young and old populations of the LMC
are easily distinguished in the (B--V) and (V--R) colors.  Evolutionary
tracks of Bertelli et al. (1994) for Z=0.001, age=8 Gyr
and Z=0.008, age=0.5 Gyr populations are overlaid on the (B-V)
color-magnitude diagram for reference.

The {\it HST} observations occurred between orbital phase 0.30 and
0.36, a flat region between the primary and secondary minima, according
to the light curve and ephemeris of Schmidtke et al. (1993).  Our V and R
magnitudes agree with out-of-eclipse magnitudes of Cowley et al.
(1991) and Schmidtke et al. (1993).  The (B--V) and (V--R) colors of
CAL~87 place it among other ordinary LMC main sequence stars of
M$_V$\squig 0, but the (U--B) color clearly sets this object apart as
something unusual.

While Companions A and B have similar U magnitudes, their (B--V) and
(V--R) colors differ considerably.  Their respective locations in the
color-magnitude diagram suggest that Companion A is an old-population
red giant, while Companion B is still on the main sequence.  Our colors
indicate that Companion A is a mid-G type subgiant.  The colors for
Companion B are not easily matched to those of any normal star, but the
photometric errors are sufficiently large that late-A star colors fit
within 2$\sigma$ error bars.  There is no obvious motivation to suspect
that A or B are related to the X-ray source.  Reference stars 1 and 3
from Cowley et al. (1990) also appear in the PC field of view and
appear high on the red giant branch in the CM diagram.  For the sake of
completeness, we provide in Table 1 the {\it HST} magnitudes for these
objects as well.

The contamination by Companions A and B to the group when CAL~87 is out
of eclipse is small but not negligible: CAL~87 contributes 80\% and
75\% of the total light of the group in V and R respectively.  However,
during eclipse, there is very significant contamination of the light
from CAL~87 by both companions.  As we have only uneclipsed observations
from HST, we must combine our maximum light, uncontaminated photometry
of all 3 stars with published reports of the eclipse depths to infer
the uncontaminated CAL~87 eclipse depth as a function of color.  This
in turn is complicated by a variety of conflicting reports in the
literature for the observed ``CAL~87" eclipse depth; some but not all
of the observers attempt to correct these values for contamination by
Companion A (past observers have been unaware of Companion B).  We use
the eclipse depths $\Delta$B=1.32, $\Delta$V=1.20, and $\Delta$R=0.99
from Cowley et al. (1990), derived from measurements which treated the
group as a single source, rather than relying on attempts from
ground-based data to deconvolve the components.  These assumptions then
permit an inferred decomposition of the 3 stars at minimum light, to
complement our direct observation of the 3 objects at CAL~87 maximum
light.  These results are shown in Table 2.  In addition to results for
each star separately and their sum, we also display results for CAL~87
+ B, as these objects cannot easily be separated in ground-based data.
The table also provides our best inference from the decomposition for
the expected uncontaminated CAL~87 eclipse depths, for use in future
modeling.

The data of Table 2 confirm the suspicion of Cowley et al. (1990) and
Cowley et al. (1991) that their optical and near IR spectra are heavily
contaminated by sources not in the CAL~87 system. Our observed colors
indicate that Companions A and B are early-G and late-A stars,
respectively; at least star A almost surely exhibits many absorption
lines in common with those that Cowley et al. (1990) report in their
blended spectrum of all 3 objects near CAL~87. Thus the explanation by
Cowley et al. (1991) that the failure of their ground-based
observations of the Ca IR triplet to reveal radial velocity variations
is due to contamination is probably correct, due not only to the
interference of continuum from the companions, but also due to the
appearance of these same spectral features in these nearby stars.

A perhaps more intriguing issue is whether the past estimates of the
spectral type of the nondegenerate star in CAL~87 are also affected by
the two contaminating companions. The inference that CAL~87 contains a
late-F or early-G star is due to the detection of Ca H \& K and the
G-band by Cowley et al. (1990) in minimum light ground-based spectra,
which we now understand to in fact contain 40\% contamination at these
wavelengths by two unrelated early-G and late-A stars, at least one of
which contains these same spectral features.  Furthermore, it is likely
that some, perhaps even most, of the remaining 60\% of the B light
comes from an incompletely eclipsed accretion disk.  Some of the
absorptions seen in the ground-based spectra may come from the CAL~87
secondary, but a careful reconsideration of this issue may now be
warranted, especially because the inference that the unseen X-ray
object is a black hole depends crucially on the inferred spectral type
of this star.  Van den Heuvel et al. (1992) presciently predicted that
the observed absorption spectrum might come from a superposed
companion, and we now at least partially verify this conjecture.

\subsection{NP Ser}

GX17+2 (=X1813-140) is a classic LMXRB (radio emission, X-ray bursts,
``Z-source" X-ray spectrum, quasi-periodic X-ray oscillations, etc.)
that was optically identified more than twenty years ago (Tarenghi \&
Reina 1972; Davidsen et al. 1976) with a $V$$\sim$$17.5$ G~star, now
known as NP~Ser, on the basis of an excellent X-ray position, and
subsequently a sub-arcsecond radio position (Hjellming 1978).
There is one problem, however: the optical ``counterpart" stubbornly
refuses to show any obvious and repeatable photometric or spectroscopic
abnormalities ({\it e.g.,} Margon 1978), despite the estimate of
$L_x/L_{opt}\sim3000$ (Bradt \& McClintock 1983).  Imamura et al.
(1987) reported a single, odd optical spike of duration 3 minutes in a
10$''$ aperture around this object, an observation to our knowledge not
replicated or confirmed in the last decade.  Naylor et al. (1991)
report possible IR variability, and colors inconsistent with a single,
normal star.  They also suggest NP Ser may be a superposition on the
X-ray source, based on incompatible absorption inferred for the optical
and X-ray objects.  Penninx et al. (1988) are of the opinion that there
is no plausible optical counterpart.

At $b=1^\circ$ and a distance of many kpc, NP Ser is a clear candidate
for an accidental superposition.  Despite the high angular resolution
achieved with these PC observations, NP~Ser appears in our data to be a
single object, with no significant deviations from the typical WFPC2
PSF of \decsec{0}{074} FWHM.  By subtracting a Tiny TIM model PSF, we
set an upper limit on a possible second nearby source which might be
the real optical counterpart as follows:  for $r>$\decsec{0}{4},
$R>23.5$; \decsec{0}{15}$>r>$\decsec{0}{4}, $R>22.5$;
\decsec{0}{04}$>r>$\decsec{0}{15}, $\Delta R>30\times r$ where $\Delta
R$ is the magnitude difference between NP~Ser and a possible
counterpart.  Thus if Naylor et al. (1991) are correct regarding a chance
superposition, the alignment may be extremely precise; however, see our
astrometric comments below.

Accurate X-ray coordinates (4$''$ radius for 90\% confidence error
circle) for GX 17+2 are available from reprocessed {\it Einstein} HRI
data in the HRICFA database obtained through the High Energy
Astrophysics Science Archive Research Center Online Service, provided
by the NASA-Goddard Space Flight Center.  Grindlay \& Seaquist (1986)
used the VLA to determine \decsec{0}{1} accuracy coordinates for a 6 cm
radio source within the {\it Einstein} error circle.  The original
optical position of Tarenghi \& Reina (1972) agrees with the radio
coordinates to within \decsec{0}{5}.  However, after an independent
positional measurement of NP~Ser using the image data and
astrometric solutions used to generate the {\it HST} Guide Star Catalog (GSC)
(Lasker et al. 1990), we find a larger-than-expected discrepancy
(2$''$) between the optical and radio positions.  Table 3 lists the published
coordinates mentioned above as well as our GSSS position.

In an investigation of the accuracy of GSC astrometry, Russell et al.
(1990) compare the difference between 48 compact radio source coordinates 
and their known optical counterpart positions measured on the GSC source
plates.  They find that the differences have $\sigma_\alpha$=\decsec{0}{63},
$\sigma_\delta$=\decsec{0}{58}, although they do find that
two of the 48 sources have a deviation of 2$''$.  Furthermore, GX 17+2
is well inside the central 50\% plate area region where other astrometric
comparisons in Russell et al. (1990) have $\sigma_\alpha$=\decsec{0}{55},
$\sigma_\delta$=\decsec{0}{53}.

Using an extraction from the digitized plates used to generate the {\it
HST} GSC (Lasker et al. 1990), we have transferred the astrometric
solution in that image to a deeper R CCD image, kindly provided by M.
Shara.  The transfer and measurement errors on the CCD are negligible
compared with systematic errors in the GSSS astrometry.  Although the
position in Tarenghi \& Reina (1972) agrees with the radio coordinates
to within \decsec{0}{5}, and Grindlay \& Seaquist (1986) cite a private
communication photographic determination which claims agreement to
within \decsec{0}{5}, we find a $\sim2.5\sigma$ discrepancy in {\it
each} coordinate $\alpha$ and $\delta$ ($\sim3.5\sigma$ total), which,
while not alarming, is worthy of note, particularly since NP~Ser
refuses to show any photometric or spectroscopic peculiarities.  A
fundamental redetermination of the radio and optical positions seems
worthwhile, and perhaps might yield a surprise.

\section{CONCLUSION}

Our imagery of V1727~Cyg reveals no close companions, and our
photometry together with an FOS spectrum in the 2300-3300 \AA\ range is
consistent with an F8 IV type star reddened by E(B--V)\squig 0.3--0.4.
Absorption line depths are similar to the IUE comparison spectrum,
suggesting that most of the light is coming from the F8 star.  For
CAL~87 we provide photometry for two close companions which contribute
up to 50\% of the flux during eclipse. Past inferences on the spectral
type of the nondegenerate star, and thus the system masses, may be
influenced by this contamination. For NP~Ser, no additional objects
which might be related to the X-ray source are revealed, but we point
out a $\sim3.5\sigma$ positional discrepancy between the radio
coordinates and the optical coordinates measured with the {\it HST}
Guide Star Catalog source data.

\acknowledgments

We thank Ralph Bohlin for his discussions about scattered light in the
FOS and help with the CALFOS software.  We are grateful to Mike Shara,
Debra Wallace, and David Zurek for providing us with the CCD image of
NP Ser, and to Christian Ready for his help at ST\,ScI.  Support for
this work was provided by NASA through grant NAG5-1630, and grant
number GO-06135.01-94A from the Space Telescope Science Institute,
which is operated by the Association of Universities for Research in
Astronomy, Incorporated, under NASA contract NAS5-26555.

\clearpage
\begin{table}
Table 1.  Photometry \\

\begin{tabular}{llllllrll} \tableline \tableline
Object  &  Reference               & \multicolumn{1}{c}{U} & \multicolumn{1}{c}{B} & \multicolumn{1}{c}{V} & \multicolumn{1}{c}{R} &  U--B  &  B--V  &  V--R  \\ \hline
V1727~Cyg  &  This work ({\it HST} WFPC2)           &  19.12  &  18.78  &  17.90  &  17.27   & 0.34  &  0.88  &  0.62  \\
   &  Cowley \& Schmidtke (1990)                    &         &  18.85  &  17.91  &          &       &  0.95  &        \\
   &  Chevalier et al. (1989)                       &         &  18.88  &  17.96  &  17.22   &       &  0.96  &  0.74  \\
   &            Kaluzny (1988)                      &  18.98  &  18.90  &  17.94  &  17.08   & 0.08  &  0.96  &  0.86  \\ 
   &  Thorstensen et al. (1979)                     &         &  18.81  &  17.88  &          &       &  0.93  &        \\ \tableline
CAL 87  &  This work ({\it HST} WFPC2)              &  17.99  &  18.94  &  18.87  &  18.73  &--0.95  &  0.07  &  0.14  \\
(maximum        &  Callanan et al. (1989) \tablenotemark{a}      &         &  18.9   &  19.05  &         &        & -0.15  &        \\
light)          &  Cowley et al. (1990) \tablenotemark{a}        &  18.30  &  19.04  &  18.90  &  18.78  &--0.74  &  0.14  &  0.12  \\
 & & & & & & & & \\
Companion A  &  This work ({\it HST} WFPC2)         &  21.93  &  21.74  &  20.90  &  20.33  &  0.19  &  0.84  &  0.57  \\
           &                                        &  (0.36) &  (0.08) &  (0.04) &  (0.03) & (0.37) & (0.09) & (0.05)  \\
           & Cowley et al. (1990)                   &         &  22.0   &  21.0   &  20.5   &        &  1.0   &  0.5   \\
           & Cowley et al. (1991)                   &         &  21.5   &  20.8   &  20.2   &        &  0.7   &  0.6   \\
 & & & & & & & & \\
Companion B  &  This work ({\it HST} WFPC2)         &  21.91  &  21.89  &  21.81  &  21.46  &  0.02  &  0.08  &  0.35  \\
           &                                        &  (0.25) &  (0.08) &  (0.07) &  (0.06) & (0.26) & (0.11) & (0.09)  \\
 & & & & & & & & \\
Ref. star 1  &  This work ({\it HST} WFPC2)         &  19.46  &  18.64  &  17.59  &  16.96  &  0.82  &  1.05  &  0.63  \\
           & Cowley et al. (1990)                   &  19.61  &  18.67  &  17.51  &  16.93  &  0.94  &  1.16  &  0.58  \\
 & & & & & & & & \\
Ref. star 3  &  This work ({\it HST} WFPC2)         &  18.88  &  18.30  &  17.37  &  16.81  &  0.58  &  0.93  &  0.55  \\
           & Cowley et al. (1990)                   &  18.93  &  18.26  &  17.27  &  16.76  &  0.67  &  0.99  &  0.51  \\ \tableline
NP Ser  &  This work ({\it HST} WFPC2)              &  19.50  &  18.63  &  17.42  &  16.65  &  0.87  &  1.21  &  0.77  \\
        &  Margon (1978)                            &  19.80  &  18.77  &  17.51  &         &  1.03  &  1.26  &        \\
\tableline
\end{tabular}
\vspace{.2in}
\tablenotetext{a}{ Includes light from Companions A and B}
\end{table}

\clearpage
\begin{table}
Table 2.  Decomposition of the Light in the CAL 87 Region \\

\begin{tabular}{lllll} \tableline \tableline
Object  &  \multicolumn{1}{c}{U} & \multicolumn{1}{c}{B} & \multicolumn{1}{c}{V} & \ \ R  \\ \tableline
    \multicolumn{5}{c}{Out of Eclipse Magnitudes} \\
CAL 87          &  17.99  &  18.94  &  18.87  &  18.73 \\
A               &  21.93  &  21.74  &  20.90  &  20.33 \\
B               &  21.91  &  21.89  &  21.81  &  21.46 \\
CAL 87 + B      &  17.96  &  18.87  &  18.80  &  18.66 \\
CAL 87 + A + B  &  17.93  &  18.80  &  18.65  &  18.44 \\ \tableline
    \multicolumn{5}{c}{Eclipse Minimum Magnitudes} \\
CAL 87          &  \ . .  &  20.71  &  20.71  &  20.39 \\
A               &  \ . .  &  21.74  &  20.90  &  20.33 \\
B               &  \ . .  &  21.89  &  21.81  &  21.46 \\
CAL 87 + B      &  \ . .  &  20.39  &  20.37  &  20.05 \\
CAL 87 + A + B  &  \ . .  &  20.12  &  19.85  &  19.43 \\ \tableline
    \multicolumn{5}{c}{Fraction of Total Light (CAL 87 + A + B) Out of Eclipse} \\
CAL 87          &   0.95  &   0.88  &   0.82  &   0.76 \\
A               &   0.03  &   0.07  &   0.13  &   0.17 \\
B               &   0.03  &   0.06  &   0.05  &   0.06 \\ \tableline
    \multicolumn{5}{c}{Fraction of Total Light (CAL 87 + A + B) at Eclipse Minimum} \\
CAL 87          &  \ . .  &   0.58  &   0.45  &   0.41 \\
A               &  \ . .  &   0.22  &   0.38  &   0.44 \\
B               &  \ . .  &   0.20  &   0.16  &   0.15 \\ \tableline
    \multicolumn{5}{c}{Inferred Uncontaminated Eclipse Depths (mag)} \\
CAL 87          &  \ . .  &   1.52  &   1.57  &   1.40 \\
\tableline
\end{tabular}
\end{table}

\clearpage
\begin{table}
\begin{center} {Table 3.  Available Positions for NP~Ser} \end{center}

\begin{tabular}{lcccccc} \tableline \tableline
Source        & $\alpha$(2000) & $\delta$(2000) &  Stated  & $\Delta\alpha$  & $\Delta\delta$  & $\Delta$tot  \\
              & 18${\rm ^h16^m}$  & $-14^\circ$02$'$  &  Error   &  from radio  &  from radio  &  from radio  \\ \tableline

VLA 6 cm \tablenotemark{a}  & \decsectim{1}{334}  & \decsec{10}{69} & \decsec{0}{1} & --              & --             & --             \\
X-ray    \tablenotemark{b}  & \decsectim{1}{2}    & \decsec{11}{15} & 4$''$         & \decsec{1}{95}  & \decsec{0}{46} & \decsec{2}{00} \\
optical  \tablenotemark{c}  & \decsectim{1}{306}  & \decsec{10}{88} & \decsec{0}{6} & \decsec{0}{41}  & \decsec{0}{19} & \decsec{0}{45} \\
optical  \tablenotemark{d}  & \decsectim{1}{439}  & \decsec{11}{92} & \decsec{0}{55}& \decsec{-1}{53} & \decsec{1}{23} & \decsec{1}{96} \\
\tableline
\end{tabular}
\tablenotetext{a}{ Grindlay \& Seaquist (1986)}
\tablenotetext{b}{ HRICFA Database at HEASARC}
\tablenotetext{c}{ Tarenghi \& Reina (1972)}
\tablenotetext{d}{ This work using CCD and GSSS images}
\end{table}


\clearpage

\begin{figure}
\plotone{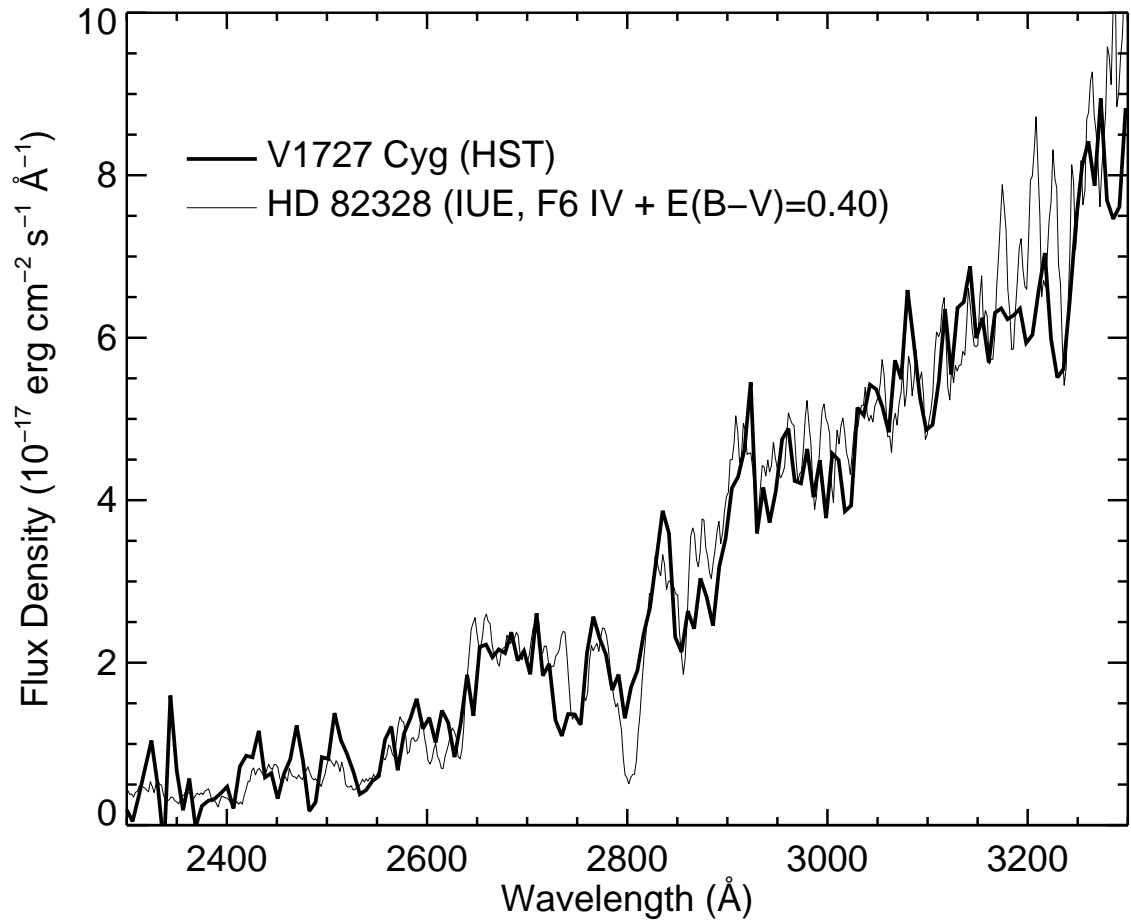}
\caption{{\it HST} spectrum of V1727 Cyg and an IUE comparison spectrum
(LWR 15500) of an F6 IV star (HD 82328) which has been reddened by
E(B--V)=0.40 and scaled to the V magnitude of V1727~Cyg.}
\end{figure}

\begin{figure}
\plotone{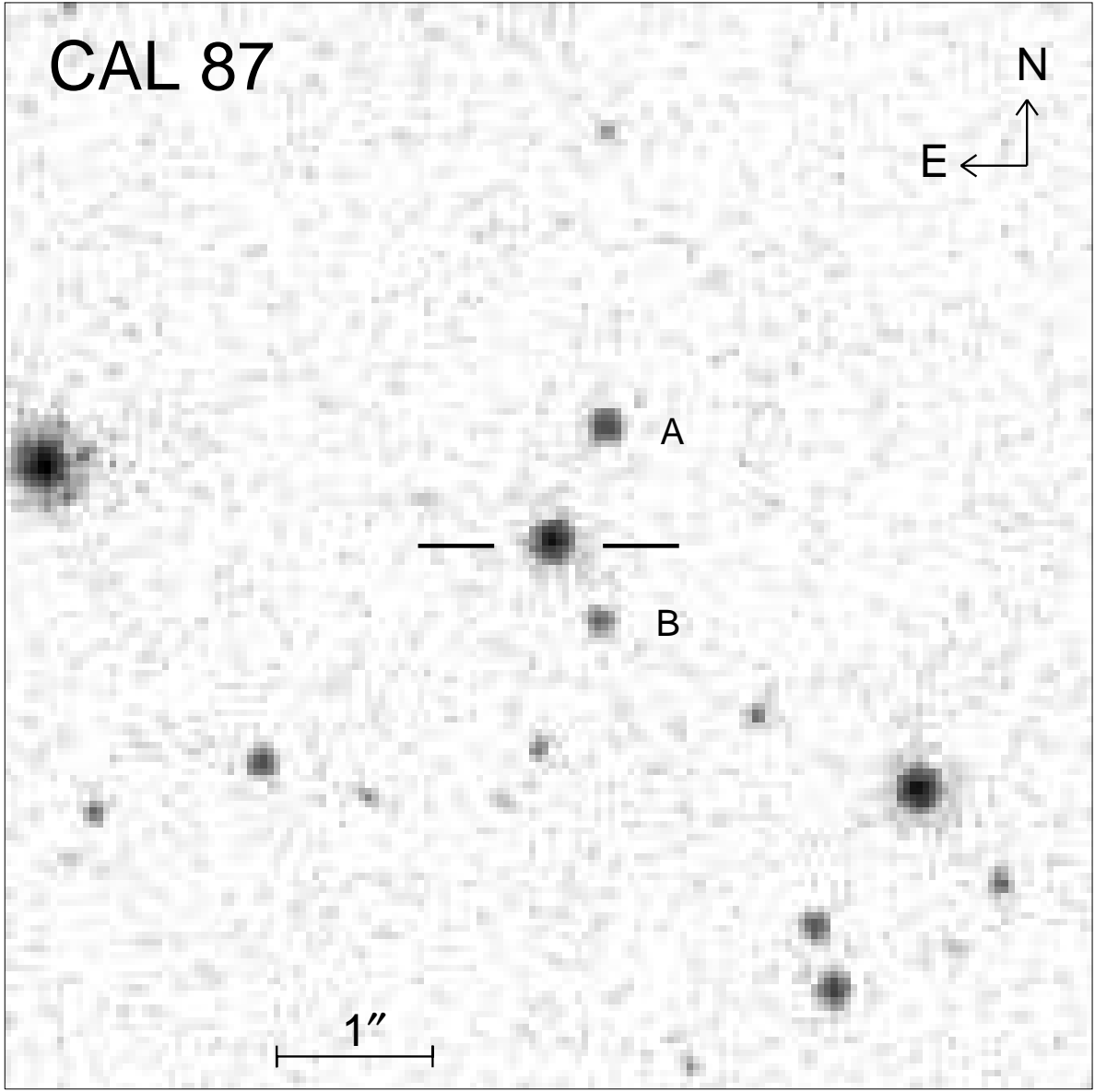}
\caption{{\it HST} WFPC2 image of CAL 87 and the surrounding
$7''\times7''$ region in the F675W filter.  CAL~87 is in the center;
close Companions A (Cowley et al. 1991) and B (this work) are separated
from CAL~87 by \decsec{0}{88} and \decsec{0}{65} respectively.  Our
photometry argues against A or B having any association with the X-ray
source.}
\end{figure}

\begin{figure}
\plotone{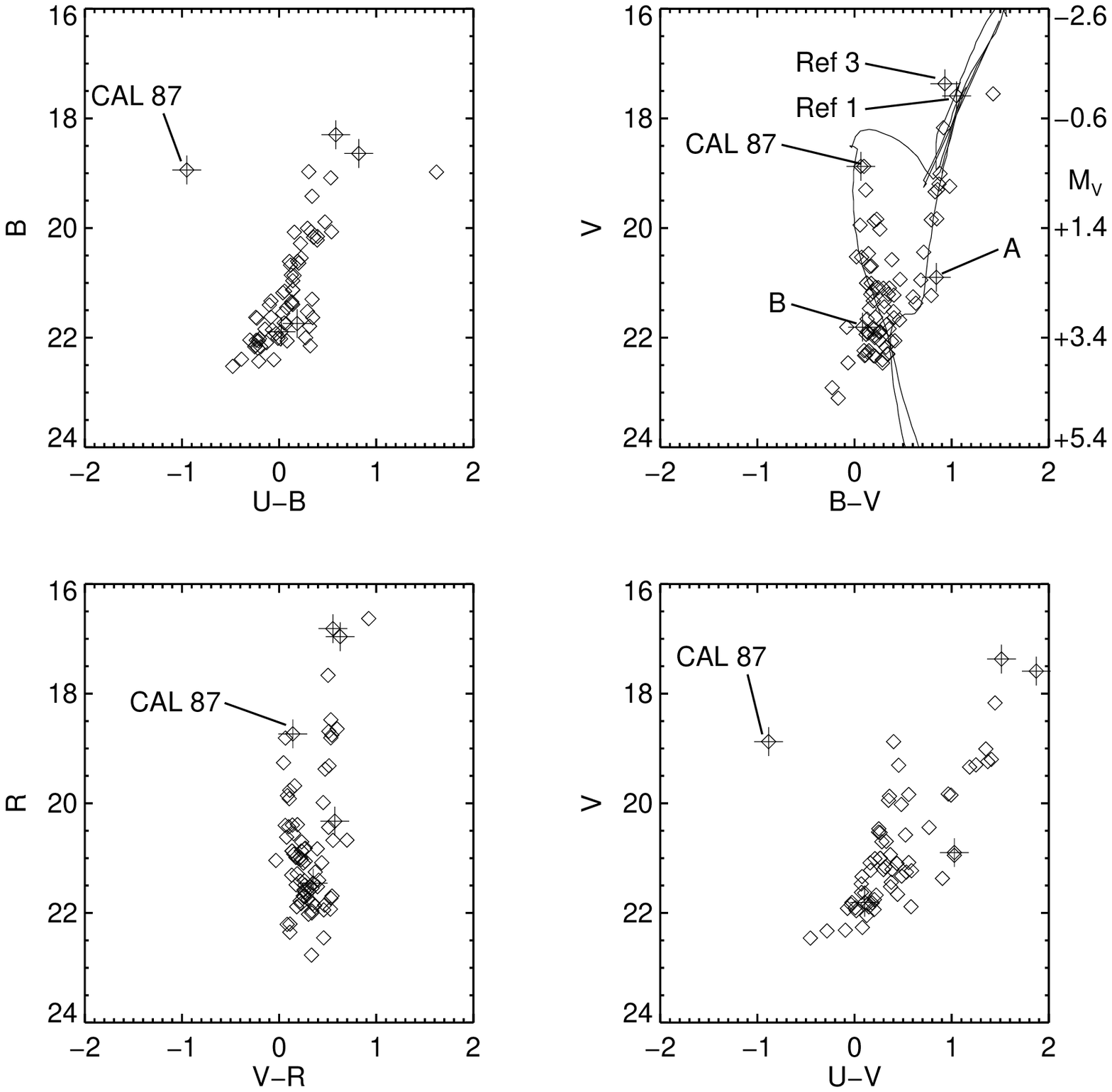}
\caption{Color-magnitude diagrams in four colors for objects on the PC
chip near CAL~87.  Objects discussed in the text are labelled.
Evolutionary tracks, detailed in the text, are drawn for reference.
The M$_{\rm V}$ scale assumes (m--M)=18.4 and A$_{\rm V}$\squig0.2.  Note
the distinct position of the X-ray source in the (U--B) diagram.}
\end{figure}


\clearpage
\begin{thebibliography}{}

\bibitem[]{} Argue, A. N., \& Sullivan, C. 1982, Observatory, 102, 4

\bibitem[Argue et al.,\ 1984]{arg84} Argue, A. N., Jauncey, D. L.,
Morabito, D. D., \& Preston, R. A. 1984, \mnras, 209, 11p

\bibitem[]{} Bertelli, G., Bressan, A., Chiosi, C., Fagotto, F. \& Nasi, E.
1994, \aaps, 106, 275

\bibitem[Bradt \& McClintock 1983]{bra83} Bradt, H. V. \& McClintock,
J. E. 1983, ARAA, 21, 63

\bibitem[]{} Callanan P. J., Machin, G., Naylor, T., \& Charles, P. A.
1989, MNRAS, 241, 37p

\bibitem[]{} Chevalier, C., 1989, Proc. 23rd ESLAB Symp. (ESA SP-296), p. 341

\bibitem[Chevalier et al.,\ 1989]{che89} Chevalier, C., Ilovaisky, S.
A., Motch, C., Pakull, M., \& Mouchet, M. 1989, \aap, 217, 108

\bibitem[]{} Cowley, A. P., \& Schmidtke, P. C. 1990, \aj, 99, 678

\bibitem[]{} Cowley, A. P., Schmidtke, P. C., Crampton, D., \&
Hutchings, J. B. 1990, \apj, 350, 288

\bibitem[]{} Cowley, A. P., Schmidtke, P. C., Crampton, D., Hutchings,
J. B., \& Bolte, M. 1991, \apj, 373, 228

\bibitem[]{} Davidsen, A., Malina, R., \& Bowyer, S. 1976, \apj, 203, 448

\bibitem[]{} Duncan, A. R., Stewart, R. T., \& Haynes, R. F. 1993,
\mnras, 265, 157

\bibitem[]{} Garcia, M., \& Callanan, P. 1995, BAAS, 27, 1323

\bibitem[]{} Garcia, M. R., \& Grindlay, J. E. 1987, \apjl, 313, L59

\bibitem[]{} Garcia, M. R., Bailyn, C. D., Grindlay, J. E., \& Molnar,
L. A. 1989, \apj, 341, L75

\bibitem[]{} Grindlay, J. E., \& Seaquist, E. R. 1986, \apj, 310, 172

\bibitem[]{} Hjellming, R. M. 1978, \apj, 221, 225

\bibitem[]{} Holtzman, J. A., Hester, J. J., Casertano, S., Trauger, J. T.,
Watson, A. M., \& The WFPC2 IDT 1995a, PASP, 107, 156

\bibitem[]{} Holtzman, J. A., Burrows, C. J., Casertano, S., Hester, J. J., 
Trauger, J. T., Watson, A. M., \& Worthey, G. 1995b, PASP, 107, 1065

\bibitem[]{} Horne, K., Verbunt, F., \& Schneider, D. P. 1986, \mnras, 218, 63

\bibitem[]{} Hughes, J. P. 1994, \apj, 427, L25

\bibitem[]{} Hutchings, J. B., Cowley, A. P., Schmidtke, P. C., \& Crampton, D.
1995, \aj, 110, 2394

\bibitem[]{} Imamura, J. N., Steinman-Cameron, T. Y., \& Middleditch, J. 1987,
ApJ, 320, L41

\bibitem[]{} Kaluzny, J. 1988, Acta Astr., 38, 207

\bibitem[]{} Kinney, A. L. 1993, in {\it Calibrating Hubble Space Telescope},
ed. J. C. Blades \& S. J. Osmer (Baltimore: STScI), p. 184

\bibitem[]{} Kinney, A. L., \& Bohlin, R. C. 1993, FOS Instrument Science
Report, CAL/FOS-103, STScI

\bibitem[]{} Koratkar, A. 1995, in {\it Calibrating Hubble Space
Telescope:  Post Servicing Mission}, ed. A. Koratkar \& C. Leitherer
(Baltimore: STScI), p. 34

\bibitem[Krist 1993]{kri93} Krist, J. 1993, in {\it Astronomical Data
Analysis Software and Systems II}, ASP Conference Series 52, ed. R. J.
Hanisch, R. J. V. Bissenden, \& J. Barnes, p. 530

\bibitem[Landsman 1993]{lan93} Landsman, W. B. 1993, in {\it Astronomical
Data Analysis Software and Systems II}, ASP Conference Series 52,
ed. R. J. Hanisch, R. J. V. Bissenden, \& J. Barnes, p. 256

\bibitem[Lasker et al.,\ 1990]{las90} Lasker, B. M., Sturch, C. R.,
McLean, B. J., Russell, J. L., Jenkner, H., \& Shara, M. M. 1990, \aj,
99, 2019

\bibitem[]{} Lindler, D. J., \& Bohlin, R. C. 1994, FOS Instrument Science
Report, CAL/FOS-125, STScI

\bibitem[]{} Long, K. S., Helfand, D. J., \& Grabelsky, D. A. 1981,
ApJ, 248, 925

\bibitem[]{} Margon, B. 1978, ApJ, 219, 613

\bibitem[]{} McClintock, J. E., Remillard, R. A., \& Margon, B. 1981,
ApJ, 243, 900

\bibitem[]{} Moneti, A. 1992, \aap, 260, L7

\bibitem[]{} Motch, C., Hasinger, G., \& Pietsch, W. 1994, \aap 284, 827

\bibitem[]{} Naylor, T., Charles, P. A., \& Longmore, A. J. 1991 MNRAS,
252, 203

\bibitem[]{} Pakull, M. W., Beuermann, K., Angebault, L. P., \&
Bianchi, L. 1987, Ap. Sp. Sci., 131, 689

\bibitem[]{} Pakull, M. W., Beuermann, K., van der Klis, M., \& van
Paradijs, J. 1988, \aap, 203, L27

\bibitem[]{} Penninx, W., Lewin, W. H. G., Zijlstra, A. A., Mitsuda, K.,
van Paradijs, J., \& van der Klis, M. 1988, Nature, 336, 146

\bibitem[]{} Pietsch, W., Steinle, H., Gottwall, M., \& Graser, U.
1986, \aap, 157, 23

\bibitem[Russell et al.,\ 1990]{rus90} Russell, J. L., Lasker, B. M.,
McLean, B. J., Sturch, C. R., \& Jenkner H. 1990, \aj, 99, 2059

\bibitem[]{} Savage, B. D. \& Mathis, J. S. 1979, \araa, 17, 73

\bibitem[]{} Schmidtke, P. C., McGrath, T. K., Cowley, A. P., \&
Frattare, L. M. 1993, \pasp, 105, 863

\bibitem[]{} Tarenghi, M. \& Reina, C. 1972, Nat. Phys. Sci. 240, 53

\bibitem[]{} Thorstensen, J. R., Charles, P., Bowyer, S., Briel, U. G.,
Doxsey, R. E., Griffiths, R. E., \& Schwartz, D. A. 1979, ApJ, 233, L57

\bibitem[]{} Thorstensen, J. R., \& Charles, P. A. 1982, ApJ, 253, 756

\bibitem[]{} Thorstensen, J. R., Brownsberger, K. R., Mook, D. E.,
Remillard, R. A., McClintock, J. E., Koo, D. C., \& Charles, P. A.
1988, ApJ, 334, 430

\bibitem[]{} van den Heuvel, E. P. J., Bhattacharya, D., Nomoto, K.,
\& Rappaport, S. A. 1992, \aap, 262, 97

\bibitem[]{} van Teeseling, A., Heise, J., \& Kahabka 1996, in IAU
Symp. 165, Compact Stars in Binaries, ed. J. van Paradijs, E. P. J. van
den Heuvel, \& E. Kuulkers (Dordrecht: Kluwer), p. 445

\bibitem[]{} Whelan, J. A., Mayo, S. K., Wickramasinghe, D. T., Murdin,
P. G., Peterson, B. A., Hawarden, T. G., Longmore, A. J., Haynes, R.
F., Goss, W. M., Simons, L. W., Caswell, J. L., Little, A. G., \&
McAdam, W. B. 1977, MNRAS, 181, 259

\bibitem[]{} Yungelson, L., Livio, M., Truran, J. W., Tutukov, A., \&
Fedorova, A. 1996 ApJ, 466, in press


\end{thebibliography}
\end{document}